\documentclass[sigconf]{acmart}

\usepackage{booktabs} 
\usepackage[english]{babel}
\usepackage[T1]{fontenc}
\usepackage{amsmath}
\usepackage{graphicx}
\usepackage{hyperref}
\usepackage{url}
\usepackage{csvsimple} 
\usepackage{caption} 
\usepackage{capt-of} 
\usepackage{algorithm}
\usepackage{hyperref}
\usepackage[noend]{algpseudocode}
\usepackage{amsmath}
\usepackage{multicol}

\setcopyright{rightsretained}

\acmDOI{}

\acmISBN{}

\acmConference[]{https://asya.ai}{\today}{Riga, Latvia}
\acmYear{2018}
\copyrightyear{2018}

\acmArticle{4}
\acmPrice{15.00}

\begin{document}

\title{asya: Mindful verbal communication using deep learning}
\subtitle{Pre-print Version, \today}

\author{Evalds Urtans}
\affiliation{%
  \institution{Riga Technical University}
  \streetaddress{Kalku iela 1}
  \city{Riga}
  \state{Latvia}
  \postcode{LV 1658}
}
\email{evalds@asya.ai}

\author{Ariel Tabaks}
\affiliation{%
    \institution{University of Central Lancashire}
   \streetaddress{Fylde Rd}
  \city{Preston, Lancashire}
  \state{UK}
  \postcode{PR1 2HE}
}
\email{ariel@asya.ai}

\begin{abstract}
asya is a mobile application that consists of deep learning models which analyze spectra of a human voice and do noise detection, speaker diarization, gender detection, tempo estimation, and classification of emotions using only voice.
All models are language agnostic and capable of running in real-time. Our speaker diarization models have accuracy over $95\%$ on the test data set. These models can be applied for a variety of areas like customer service improvement, sales effective conversations, psychology and couples therapy.
\end{abstract}

%
%



\maketitle


\section{Introduction}
\par
asya is a mobile application that listens to a person's voice and provides private feedback on a person's verbal communication. It  gives metrics on how much a person listens and speaks, how long are a person's sentences and utterances, how fast a person speaks, how positive is the tone of a person, the confidence level of a person’s voice based on the tone and other metrics. It is deployed as a web service and stand-alone application. These models are being applied to a variety of tasks starting from customer service evaluation to analysis of private conversations and couple's relationships therapy.
\par
Neuroscientists Andrew Newberg, M.D., and Mark Waldman, have identified through brain scans and from other studies that if everyday verbal interactions is coupled with increased moment-to-moment awareness the results can lead to increased levels of trust building, resolved conflicts, increased intimacy and other benefits \cite{andrewnewbergm.dWordsCanChange2014}. 
The findings show that people can benefit from speaking less, shorter, and slower as human brain short-term memory holds only about four “chunks” of information, which translates to speaking time under 30 seconds \cite{andrewnewbergm.dWordsCanChange2014}. Furthermore, when people practice the 30 second rule, they can train themselves to increase awareness of filtering out lower quality information. In addition, when speaking is kept with brevity in mind, the emotional centers of the brain that can be triggered by certain words are less likely to lead the speaker to sabotage the conversation. The research also shows that exercising awareness not only helps to increase connection with other people, but also suppresses the brain’s ability to generate feelings of anxiety, irritability or stress. 
\par
The tone of voice, emotions, and the way how a person speaks are as important as the content that a person speaks. For example, one could imagine how one could say the phrase "you are such a fool" in a way that could offend someone or in a way that could be even playful and funny.

\section{Related work}
\par
In the last decade, there has been great progress in natural language processing (NLP) models to do the tasks like "text-to-speech", "speech-to-text", translation, semantic and syntactic understanding of language
 \cite{DBLP:conf/ssw/OordDZSVGKSK16} \cite{DBLP:conf/interspeech/ZhouHKVD18} \cite{DBLP:journals/corr/abs-1711-02132} \cite{DBLP:journals/corr/abs-1711-11293}.
However, only in recent years there has been emerging research to use deep learning models to analyze human's voice biometric features that are linked to psychology \cite{DBLP:journals/corr/LiMJLZLCKZ17} \cite{DBLP:conf/icassp/Bredin17}. Before it has been done using classical machine learning models that yielded in lesser results \cite{DBLP:conf/ismir/MathieuEFPR10}.
Voice features that typically are used for analysis are Mel-frequency filter banks, Log Mel spectrograms, Mel-frequency cepstrum (MFCC), or even raw waveforms in combination with audio envelopes. Mel-frequency filter banks are filters that are  applied to spectrum calculated by Fast Fourier Transform (FFT) to simulate specific amplitudes of sounds at different frequencies that are audible to the human ear. The audio spectrum is much broader than that what human ear can perceive, but other frequencies of sound are less likely to contain useful information for NLP tasks.

\begin{figure}[H]
\includegraphics[width=0.5\columnwidth]{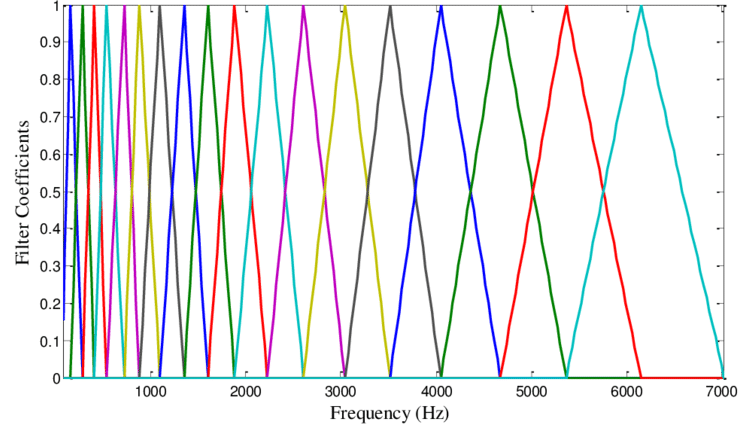}
\caption{Mel-frequencies filter bank to emulate human's ear perception.}
\end{figure}

\par
Recently, also some progress has been done into recognizing emotions form a person's voice. Historically, there have been very little datasets available for voice alone, but recent techniques using transfer learning enabled to accumulate considerable datasets with unsupervised learning to analyze emotions in a human's voice \cite{DBLP:conf/mm/AlbanieNVZ18}.
Some of the recent works have used deep Convolutional Networks (ConvNet) \cite{NIPS2012_4824} to extract features from human's voice spectra and classify 8 basic emotions: Happiness, Sadness, Anger, Fear, Disgust, Surprise, Boredom and Neutrality. Even with the basic ConvNet model, it has been possible to surpass human reference accuracy on detecting emotions in a person's voice. For example, humans on average were able to detect happiness in voice with $65\%$ precision, whereas ConvNet model was able to detect it with $100\%$ precision \cite{somayehshahsavaraniSpeechEmotionRecognition2018}, \cite{DBLP:journals/corr/NiuZNHT17}. 

\begin{figure}[H]
\includegraphics[width=1.0\columnwidth]{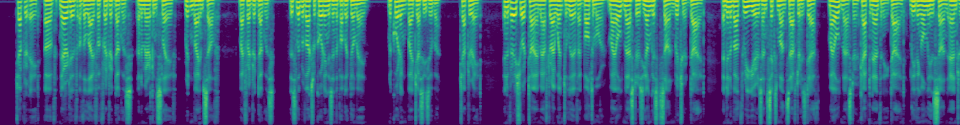}
\caption{Mel-spectogram of a person speaking (processed input of asya models).}
\end{figure}

\section{Methodology}

asya uses deep learning models that take as an input \\ Mel spectrograms and other features from raw voice recordings. Deep Residiual Networks (ResNet) \cite{DBLP:conf/cvpr/HeZRS16} and DenseNet \cite{DBLP:conf/cvpr/HuangLMW17} models have been applied as feature encoders.
ConvNet models are deep artificial neural network models that have very similar results when experimentally compared to human retina natural neural networks \cite{DBLP:journals/ploscb/KubiliusBB16}.
 At first layers of the model, they extract basic features like Gabor patches and edges, but the deeper they go they extract more general features. For example, for face detection task, first they would detect features like nose and eyes, but then in deeper layers faces as a whole.
These models are very deep with usually more than 32 layers. ConvNet models with residual connections (ResNet) allow error to flow freely using back-propagation algorithm without vanishing gradient problem. 
In case of DenseNet, there are even more connections and better flow of gradient of error through the model.
asya models have been trained on multiple large private datasets from different speakers and languages using DenseNet models and other proprietary models. 

\begin{figure}[H]
\includegraphics[width=1.0\columnwidth]{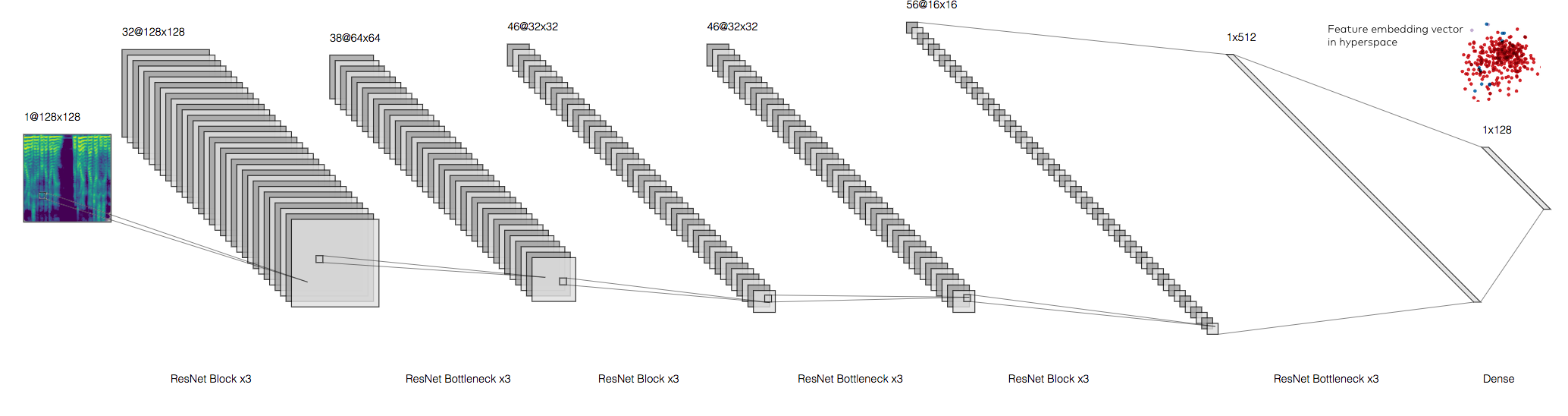}
\caption{Schematic illustration of ConvNet model for speaker feature embedding using Mel-spectogram sample.}
\end{figure}


\subsection{Speaker diarization}

Speaker diarization (identification of different speakers in parts of utterances) has been done primary using i-vector, d-vector \cite{Snyder2017DeepNN}, x-vector \cite{Snyder2018XVectorsRD} based models. 

More recently, RNN based models also have been applied like UIS-RNN \cite{Zhang2019FullySS}. 

It has also been done using triplet loss or contrastive loss and cosine similarity of embedding vectors (fingerprint vectors of human voice) \cite{DBLP:journals/corr/LiMJLZLCKZ17} \cite{DBLP:conf/icassp/Bredin17} \cite{DBLP:conf/cvpr/SchroffKP15}. asya uses Exponential Triplet Loss function and clustering of speaker embedding to achieve speaker diarization and speaker re-identification in one step \cite{Urtans2020ExponentialTL}.

asya models are utterance (phrase) and language independent, whereas, for example, Google Home recognize speaker by specific phrase like "Ok, Google". It means that asya models are capable of identifying a person's voice at any point in natural conversations.
To improve training results of speaker diariaztion models data have been split into multiple parts. asya has been trained as a set of hierarchical models that first predict if the audio in a given window is a noise or speech, then if it is a man or woman and finally does feature embedding of person's voice. During the testing, we use the center of the mass of a person's voice embedding vector to estimate a probability of voice sample belonging to a particular person \footnote{\url{https://yellowrobot.xyz/asya_demo.html}}.


\begin{figure}[H]
\includegraphics[width=1.0\columnwidth]{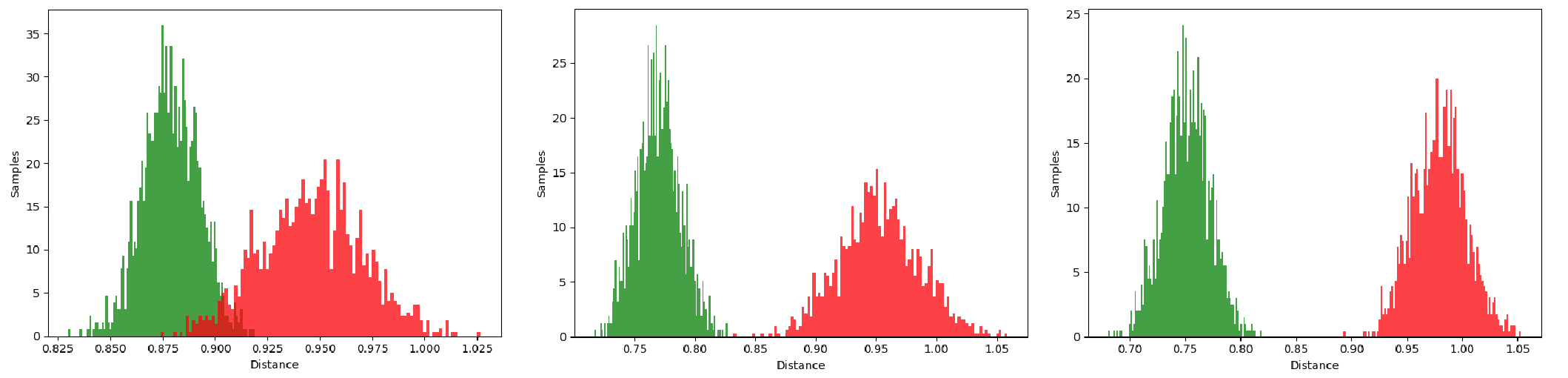}
\caption{Histograms of cosine distance between speaker embedding vectors acquired by asya male models. Left in the beginning of training, right after training have been completed. Green are distances between samples form same speaker. Red are distances between samples form different speakers. }
\end{figure}

\begin{figure}[H]
\includegraphics[width=1.0\columnwidth]{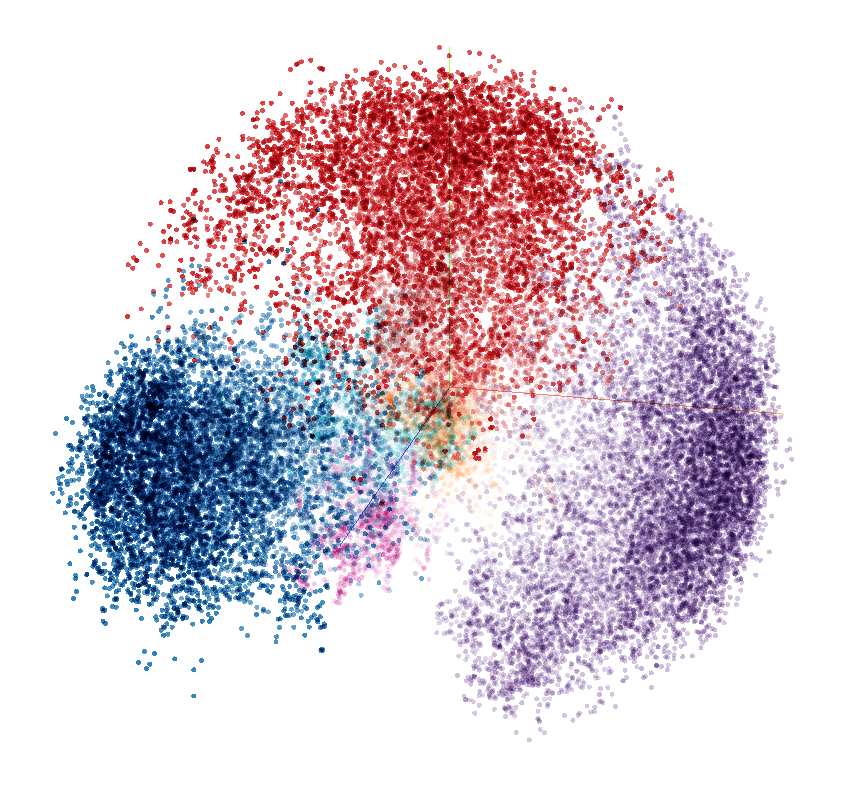}
\caption{Representation of speaker embedding vectors from asya models in 3D space (Spherical PCA). Colors denote different speaker samples in test data set.}
\end{figure}

\begin{figure}[H]
\includegraphics[width=1.0\columnwidth]{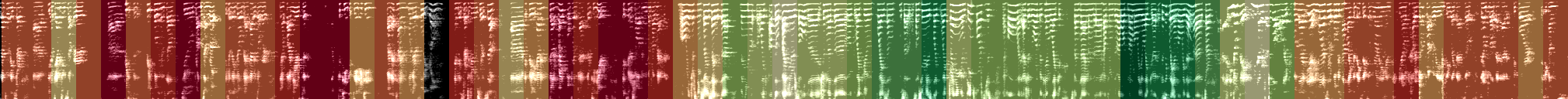}
\caption{Segmentation of audio sample for speaker diarization task for 2 male speakers. Red frames are furthest away from speaker feature embedding vector and green frames are closest thus identifying target speaker.}
\end{figure}

\subsection{Emotion classification}

Speaker emotion classification task is also done using the same deep learning feature extractor as for speaker diarization task, but with softmax loss function at the end \cite{somayehshahsavaraniSpeechEmotionRecognition2018}. 

Standard academic datasets of classified emotional states of audio are read to validate approaches. For example, traditional German EmoDB dataset \cite{Burkhardt2005ADO} contains only 500 samples of 10 speakers. 

The methodology for our work initially has been based on EmoVoxCeleb
\cite{DBLP:conf/mm/AlbanieNVZ18}. EmoVoxCeleb is trained on FERPlus \cite{BarsoumICMI2016} dataset of still pictures of human FER (Facial Expressions) in different emotional states and then applied to a larger VoxCeleb dataset of videos \cite{Nagrani2017VoxCelebAL}. These emotional states are classified as Paul Ekman's 8 basic emotions.

Even though we can achieve state-of-art results in academic data-sets we had to create our own proprietary methodology and dataset to to reach similar performance in production systems. 

To acquire more training, data transfer learning and unsupervised learning have been used to scrape public data from video sites in the internet.

\begin{table}[H]
\centering
 \captionof{table}{Preliminary results of emotion classification using asya models}
\begin{filecontents*}{./tables/emotions-1.csv}
emotion ,Human [14], EmoVoxCeleb, asya
happiness ,84, 35, 62
sadness ,81, 71, 75
anger ,97, 72, 75
fear ,84, 35, 75
surprise ,81, 36, 69
disgust ,67, 67, 75
neutral,93, N/A, 69
\end{filecontents*}
\csvautotabular{./tables/emotions-1.csv}  
\end{table}

\section{Results}

Asya models are currently in development and are being tested using mobile application in natural conversations to improve couples relationships through conversations in a similar manner, how does couple therapy would work.

Set of hierarchical models are executed in real-time in less than 500 ms. for every 1 sec. frame on consumer grade GPU server. It is also possible also to deploy and execute these models on flagship mobile phones with machine learning specialized processing units.

\begin{table}[H]
\centering
 \captionof{table}{Preliminary results of speaker re-identification and other voice classification tasks of asya models}
\begin{filecontents*}{./tables/results-1.csv}
model,accuracy
noise detection, 99.2
gender detection, 89.3
male speaker re-identification, 87.4
female speaker re-identification, 88.4
\end{filecontents*}
\csvautotabular{./tables/results-1.csv}  
\end{table}

\section{Conclusions}
\par
The proposed models are capable of analyzing human's voice in real-time. asya is able to detect noise in audio samples and process only parts with a human's voice. asya models are able to detect a human's perception of the gender of the speaker with high precision. Finally, it is able to produce a unique embedding vector for each person's voice to combine speaker diarization and reidentification tasks in a single step. Furthermore, asya models are able to detect Ekman's basic human emotions from language and utterance independent data.
\par
There can be a wide range of use cases for asya models. It has been successfully deployed to improve communication skills and encourage mindful communication in a commercial product \url{https://asya.ai}. 
Asya is being developed also to improve public speaking skills and provide feedback for psychologists about their sessions with patients.
Asya models also can be used to monitor customer experience in customer service-centered businesses like phone hotlines, post offices, stores, telemarketing, etc. 
Finally, they could be used to identify persons of interest in large databases of audio recordings, but there are even more use cases than listed in this paper.
\nocite{*}
\bibliographystyle{ACM-Reference-Format}
\bibliography{main}

\end{document}